\documentstyle[12pt]{article}
\begin{document}

\begin{titlepage}
\centerline{\large\bf INSTITUTE OF THEORETICAL AND EXPERIMENTAL PHYSICS}
\vspace{7cm}
\centerline{\large\bf A.Yu.Dubin, A.B.Kaidalov and Yu.A.Simonov}
\vspace{2cm}
\centerline{\large\bf THE QCD STRING WITH QUARKS.}
\vspace{5mm}
\centerline{\large\bf I.SPINLESS QUARKS}
\vspace{5cm}

\centerline{\large\bf Moscow - 1993}
\vspace{1cm}
$~~$

\end{titlepage}

\newpage
$~~$
\vspace{60mm}

\centerline{\large\bf A b s t r a c t}
\vspace{5mm}
\large
{}~~~~~Starting from the QCD Lagrangian we derive
the effective action for massive quark and antiquark at large  distances,
corresponding to the minimal area low of the Wilson loop.  The path integral
method is used to quantize the system and  the spectrum is obtained with
asymptotically linear Regge trajectories. Two dynamical  regimes
distinguished by the string energy--momentum distribution are found: at
large orbital excitations ($l\gg 1$) the system  behaves as a string and
yields the Regge slope of $\frac{1}{2 \pi \sigma}$,
while at small $l$ one obtains a
potential-like regime for relativistic or nonrelativistic system .
 The problem of relative time is
clarified.
It is  shown that in the valence quark approximation
 one can reduce the initial
four-dimensional dynamics to the three-dimensional one.

The limiting case of  a pure string (without quark kinetic terms)
is studied and the spectrum of the straight-line string is obtained.

\newpage
$~~$

\large

\section{Introduction}

This is the first paper of the presumed series devoted to the quantum dynamics
of the quark-antiquark system at large distances. Our starting point is the
formalism of
vacuum correlators, developed previously [1] (for a review
see [2]). It allows one to represent the gauge-invariant Green's function of
the
quark-antiquark system in a form, where all dynamics is contained in the
averaged
Wilson loop operator.

We simplify our problem by disregarding effects due to the quark spins and
additional quark
loops (sea quarks) having in mind to come back to it
in later papers.

In this way our simplified problem  is that of a scalar quark and antiquark
without
additional  quark loops in the confining background field.

One of the most important point of the paper is to identify an explicit
mechanism creating the QCD string and to find the properties of the latter.

Even the question: what is it, the QCD string? is not trivial.
Usually it  is associated with the  Nambu-Goto string  action (for review
see [3]) describing the open string with  active degrees of freedom all
along the string  including the ends, where massless quarks are presumed to
 be described by the proper boundary conditions.  This standard picture  is
 plagued by unphysical features in 4 dimensions and needs $d=26$ or
 supersymmetric extentions to be consistent [3].

 The picture which emerges in our paper and based on the QCD Lagrangian and
 the vacuum correlator method essentially differs from the standard one.
 First, the string appearing in the $q\bar{q}$ system has a world-sheet
coinciding with the
 surface appearing in the area law of the Wilson loop. This area law is a
 natural consequence of the cluster expansion for the Wilson loop [4], and
 the surface bounded by the Wilson contour appears also naturally in the
 formalism.

 An important point is what kind of surface appears in the area law of the
 Wilson loop ?~~ Since the whole cumulant series  does  not  depend on the
  shape of the surface and has  no degrees of freedom on it , we come to the
conclusion that
  this should be the minimal surface which enters the area law and
  therefore defines also the shape of the string, connecting quark and
antiquark.
  We stress however
   that this  is an assumption we make here \footnote{This assumption is
   equivalent to the uniform convergence of the cluster series for the
   Wilson loop [4], since any term in the series provides the area law for
   large enough loop.}, and it is desirable to obtain the behaviour of the
   Wilson loop in QCD directly.

    The QCD area law  appears only asymptotically at large distances,
$R\gg T_g$, where $T_g$ is the so-called gluon correlation length, which
enters as a scale for  the vacuum correlators [5,1,2]. Recent Monte-Carlo
calculations give the estimate $T_g \approx  0.2 \div  0.25 fm$ [6]  and
therefore we can apply our formalism even to the ground state mesons (see
[2] for a discussion).
   We approximate the minimal area law surface as the world sheet surface of
the straight line string connecting proper positions of quark and
antiquark.
  We shall discuss in
the next sections the validity of the straight-line approximation.

 We
consider the string tension as being renormalized and disregard all
corrections including perturbative gluon exchanges and the creation of
additional quark loops, since our aim here is to concentrate on the main
dynamical ingredient -- the interaction between quarks corresponding to the
minimal area law.

Therefore in this approximation the minimal string may rotate (orbital
excitations),  and oscillate
longitudinally (streching and expanding). The latter type of motion is not
possible for the standard Nambu-Goto string.

After these introductory words about the definition of the QCD string, we can
outline
the purpose and the plan of the paper.

We shall quantize the system  consisting of a massive or massless quark and
antiquark connected by the minimal string.

We shall extensively use the method exploited recently by one of us [2,7]
for this purpose. The method contains five steps. (1) First, one represents
the quark-antiquark Green's function using the Feynman-Schwinger form [1]
as a path integral over trajectories of $q$ and $\bar{q}$
parametrized by their proper time parameters. (2) Second, all
gluonic nonperturbative field contained in Wilson loops, is replaces
by the minimal area law. (3) Third, the minimal area is assumed to be
spanned by straight lines connecting a point on the $q$ trajectory with
another point on the $\bar{q}$  trajectory, these two points are chosen at
the same proper time parameter. (4) The proper-time Hamiltonian is introduced
to get rid of path
integration and obtain instead differential equation. (5) Fifth, an
approximation is made in the method [2,7] to expand
the Nambu-Goto string term in the action in powers of some numerical
parameter (which yields also expansion in powers of relative
velocity $\dot{r}^2$).

Due to this expansion dynamics in relative time is free and one
integrates it out, regaining in the end three dimensional
(relativistic) dynamics with dynamically generated constituent mass.

This approximation [1,2,7] appeared to be simple
and practical, yielding masses and Regge trajectories in terms of only one
parameter -- the string tension.  In this way light mesons, heavy-light
mesons [8], baryons [9] and glueballs [10] have been considered.  The
accuracy of the approximation  for masses was estimated as $\sim 10\%$.

As a result [7] the slope of Regge trajectories was obtained to be
$(8\sigma)^{-1}$
the same as in the potential models [11]
in contrast to the Nambu--Goto string slope of $(2\pi\sigma)^{-1}$.

In the present paper we abandon this approximation and treat the effective
action of the system exactly, using the  formalism of auxiliary
fields to get rid of the  square root term [12,3].
As a result we shall not only improve the accuracy of the
approximation but shall obtain a qualitatively different dynamics,
which has not been present before in the approximation [2,7].

 We shall find that there are two regimes of $q\bar{q}$
dynamics distingushed by the energy- momentum distribution of gluon
fields, --- for large orbital momenta $l$ the resulting spectrum in the
leading approximation  coincides with that of a pure string with a slope
$1/2\pi\sigma$ , while for low values of $l$ (depending on the masses of
quarks) the dynamics is described by the relativistic potential-like
approach,close to the one obtained  previously [8,2]. For the heavy quark
system the potential picture is valid in a large region of $l$ and it joins
smoothly the string picture for very large $l$.

The limiting case of pure strings (without kinetic terms of quarks) is
considered in detail
\footnote{A short version of these results is reported in ref.[13]}. The
straight-line Regge-trajectories, corresponding to the spectrum of the
rotating straight-line string is obtained. This is in an agreement  with
the results obtained in [14].  Note that in our approach the
string picture has not been assumed but it was derived from the
QCD-Lagrangian under rather general assumptions.

The important issue of
relativistic dynamical systems is the role of the relative time of $q$ and
$\bar{q}$. To make our straight-line approximation for the minimal surface
selfconsistent we have to integrate over the class of paths without time
backtracking of quarks (in the c.m.s.). This constraint corresponds to the
valence quark approximation to the problem and allows one to rigorously
reduce the initial four-dimensional (4D) dynamics to the three-dimensional
(3D) one.  As a result the quadratic 4D kinetic terms are transformed  into
3D square-root-type terms.  We have the Lorentz invariant effective action
with the constraint, which can be formulated in a Lorentz covariant form. So
the calculation of the spectrum can be  performed in an arbitrary frame
and for the sake of convenience we work in the meson rest system.

The plan of the paper is the
following.  The Feynman-Schwinger representation (FSR) for the $q\bar{q}$ -
Green's function is given and the effective action for $q\bar{q}$-system is
obtained in Sec.2.  In section 3 we obtain the approximate explicit
expression for Wilson loop at large distances.  In section 4 in the valence
quarks approximation we integrate out time-components, so that we reduce the
4D dynamics to the three-dimensional one.  Gaussian representation for the
obtained action is formulated in Sec.5.  The method of auxiliary fields
[3],[12] is used in this section to get rid of the square root term, which
 determines the string action.

The  case of a pure straight-line string without quarks at the ends  is
discussed in Section 6. In Section 7 Hamiltonian for the  general case of
the straight-line string with quarks at the ends is obtained and in two
limiting cases $l \approx 1$ and $l\gg 1$ the analytic form of the
spectrum is established.

In conclusions we summarize our main results and make comparison to those in
literature.

Appendices A,B,C contain technical  details of the derivation of formulas in
the text.

\section{Green's function of the quark-antiquark system interacting with gluon
field.}

In this Section we use the Feynman-Schwinger representation  for the
quark-antiquark Green's function
 to obtain the effective action for $q\bar{q}$
system in terms of the Wllson loop [1].

We start with the initial and final $q\bar{q}$ states defined on  space-like
surfaces in a gauge-invariant way
\begin{equation}
\Psi_{in} (y,\bar{y}) = \bar{u}(y) \Gamma_{in}(y,\bar{y}) u(\bar{y})
\end{equation}
\begin{equation}
\Psi_{out} (x,\bar{x}) = \bar{u}(x) \Gamma_{out}(x,\bar{x}) u(\bar{x})
\end{equation}
where $ \Gamma_{in},\Gamma_{out}$ contain a parallel transporter
$\Phi$ and some vertex with definite Lorentz structure
$\gamma_{in},\gamma_{out}$:
\begin{equation}
\Gamma_{in} = \Phi (y,\bar{y}) \gamma_{in} ,\;\;\Phi(y,\bar{y})= P exp ( ig
\int^y_{\bar{y}} A_{\mu} d z_{\mu})
\end{equation}
and the same for
$\Gamma_{out}; \gamma_{in}$ may be e.g.$\gamma_5,\gamma_{\mu}$ etc. The
operator $P$ ensures the  the ordering of $A_{\mu}(z)$ along the path
$z_{\mu}(t)$. The Green's function $G(x\bar{x}/y\bar{y})$ is obtained by
averaging the product $\Psi_{in} \Psi_{out}^+$ over all quark and gluonic
fields inside the path integral with the usual QCD action
\begin{equation}
G(x\bar{x}/y\bar{y}) = <\Psi_{in}(y,\bar{y})\Psi_{out}^+(x,\bar{x})>_{\Psi,
A}
\end{equation}

The quark degrees of freedom can be easily integrated out with the result
(for a nonzero flavour channel)
\begin{equation}
G(x\bar{x}/y\bar{y}) = <tr \Gamma (\bar{x},x) S(x,y)
\Gamma(y,\bar{y})S(\bar{y},\bar{x})>_A
\end{equation}
 where $S(x,y)$ is the
quark propagator in the gluonic field $A_{\mu}$, and the averaging over
 gluonic fields now includes also the quark determinant (actually a product
of determinants over all flavours):
\begin{equation}
<B>_A \equiv \int DA
e^{-S(A)} B(A)\prod_{i} det (m_i + \hat{D}(A))
\end{equation}

For the quark propagator in the gluonic field $A_{\mu}$ one can use the
Feynman-Schwinger representation [1,2,15]
\begin{eqnarray}
\nonumber
S(x,y) = <x\mid (m+\hat{D}(A))^{-1}\mid y> =\\
=<x\mid (m-\hat{D}(A)) \int^{\infty}_{0} ds e^{-s(m^2-\hat{D}^2(A))} \mid y>=\\
\nonumber
=(m-\hat{D}(A))_x \int^{\infty}_{0} ds Dz e^{-K}\Phi_{\Sigma}(x,y)
\end{eqnarray}
where the following notations are used
\begin{equation}
K= m^2s + \frac{1}{4} \int^{s}_{0} \dot{z}^2_{\mu} (t) dt ,\;\;\;\;\;
\dot{z}_{\mu}= \frac {d z_{\mu}(t)}{d t}
\end{equation}
\begin{equation}
\Phi_{\Sigma}(x,y) = (P_{\Sigma} exp( g\int^{s}_{0}\Sigma_{\mu\nu}
F_{\mu\nu} (z(t))dt))\Phi(x,y)
\end{equation}
and $\Sigma_{\mu\nu}= \frac{i}{4}(\gamma_{\mu} \gamma_{\nu}-
\gamma_{\nu}\gamma_{\mu})$.  The operators $P_{\Sigma}$ ensure
 the proper  ordered
insertion of operators $\Sigma_{\mu\nu}F_{\mu\nu}$ along this path (for a
discussion see [15]).

 Insertion of (7) into (5) immediatly gives [8,15]
\begin{eqnarray}
\nonumber
G(x\bar{x} \mid y\bar{y})=\\
 \int^{\infty}_{0} ds e^{-K}Dz \int^{\infty}_{0}
d\bar{s} e^{-\bar{K}} D\bar{z} <\gamma^+_{out} \Phi(\bar{x}, x)
(m-\hat{D}(A))_x \Phi_{\Sigma}(x,y)\times\\
\nonumber
\times
\gamma_{in}\Phi(y,\bar{y})(m-\hat{D}(A))_{\bar{y}}
\Phi_{\Sigma}(\bar{y},\bar{x})>_A
\end{eqnarray}

Analysis of the factors $(m-\hat{D})$ made in [9] shows that it can be
replaced by
$(m+\frac{1}{2}\hat{\dot{z}})$ and taken out of the angular brackets, so that
we
are left with the only term defining all the dynamics of the system, both
perturbative
and nonperturbative, which we call $W_{\Sigma}$.

The presence of spin degrees of freedom in (7) makes dynamics rather
complicated. For
this reason we shall omit the spin dependence  and concentrate
on the confining dynamics without spins.
We also neglect the additional quark loops, which corresponds to the
substitution of  $\Pi det$ in eq. (6) by 1. This is in line with usual
quenched approximation and can be justified in the $1/N_c$ expansion.

The resulting Green's function in this case has the form
\begin{equation}
 G(x\bar{x}\mid
y,\bar{y}) = \int^{\infty}_{0}ds \int^{\infty}_{0} d\bar{s} e ^{-K-\bar{K}}
Dz D\bar{z}<W(C)>_A
\end{equation}
 where $W(C)$ is the usual Wilson loop
operator
\begin{equation}
W_{\Sigma}(C)= tr \Phi
(\bar{x},x)\Phi_{\Sigma}(x,y) \Phi( y,\bar{y})
\Phi_{\Sigma}(\bar{y},\bar{x})
= tr P exp [ig \int_{C}A_{\mu} dz_{\mu}]
\end{equation}

The closed contour $C$ consists of initial and final pieces
$[x,\bar{x}],[y,\bar{y}]$
and paths $z(t), \bar{z}(\bar{t})$ of the quark and antiquark.

It is convenient  introduce as in [2] the following "center of mass"
coordinate $R_{\mu}$ and the "relative" coordinate $r_{\mu}$
$$R_{\mu} = \frac{s\bar{s}}{s+\bar{s}}(\frac{1}{s} z(t) +\frac{1}{s}
\bar{z}(t))$$
$$r_{\mu} = z(t)-\bar{z}(t)$$
with the boundary condition imposed in the Lorentz covariant way
$$ R_{\mu}(s)- R_{\mu} (0) = Tu_{\mu}\;\;\; u^{\mu}u_{\mu}=1$$
and $r_{\mu}(s)= r_{\mu}(0)$ are also fixed.

In what follows we shall be interested in the limit $T\rightarrow \infty$
to consider  asymptotical states of the system. In this limit our boundary
condition corresponds to the total momentum of the meson $P_{\mu}\sim
u_{\mu}$.

\section{Evaluation of the Wilson loop at large distances}

In this section we shall obtain approximate expression for
$<W(C)>_{A}$ at large distances.

Using the cluster expansion we can represent $<W(C)>$ )
 as [4]
\begin{equation}
<W(C)>=  exp\sum^{\infty}_{k=1} \frac{(ig)^k}{k!} \int
d\sigma(1)...d\sigma(k)\ll
F(1)...F(k)\gg
\end{equation}
where cumulants are irreducible averages [16,17] and we omit Lorenz indices
in $d\sigma_{\mu\nu}(u(i))$ and $F_{\mu\nu}(u(i))$.

As shown in [2,18], there are three possible regimes for $<W(C)>$ depending
on the relation between the sizes of the loop $C$ and the correlation length
$T_g$
, defining the  decay of cumulants $\ll F(1)...F(i)...F(j)...F(k)\gg \sim
exp (-\frac{\mid z(i)-z(j)\mid}{T_g})$.

If we represent the loop $C$ as a rectangular of time length $T$ and space
width $R$,
then we have (omitting perimeter-type terms always coming from the quark-mass
renormalization
and exchanges) [2]
'\begin{equation}
i)<W(C)>\approx exp (-\sigma S) , R\gg T_g, T\gg T_g
\end{equation}
\begin{equation}
ii)<W(C)>\approx exp (-T(c_2R^2+c_4R^4+...)), R \ll T_g, T\gg T_g
\end{equation}
\begin{equation}
iii)<W(C)>\approx exp (-\frac{S^2g^2<F^2(0)>}{24N_c}+ 0(S^3)),
R\ll T_g, T < T_g
\end{equation}

Here $S$ is the area of the surface inside the contour $C$, which we take as
the minimal
surface, since $<W(C)>$ does not depend on the shape of the surface [17].

Thus we see that the area law (14) and the ensuing string dynamics appears
only as the asymptotic regime for large contours, while for small or narrow
loops the area law (and the QCD string) is absent and is replaced by a much
weaker interaction.

In what follows we concentrate on the regime (14), i.e. we shall consider
only large loops, $R,T\gg T_g$.

We note that in the Monte Carlo calculations this law
 has been seen in the wide region of $R, 0.1 \preceq
R\preceq 1 fm $ [19] also in the presence of dynamical quarks (the
determinantal term in (6)) [20]. From those results one could conclude that
$T_g$ is small enough, so that the regime (14) is dominant for light quark
systems.

Independent Monte-Carlo calculations [6] have confirmed that $T_g \cong 0.2
\div 0.3 fm$, which should be compared with characteristic quark relative
distance and  orbiting time
$R \sim T_q\approx 1 fm$ for light quark system [2]. Thus we can conclude that
the area
law (14) is a reasonable approximation for light quark systems (even for
ground states) and for excited states of heavy quarkonia.

The "minimal area law" implies the neglect of gluonic excitations above the
QCD vacuum, which could lead to an effective integration over the surfaces,
bounded by the contour $C $.

 Next step is to construct explicitly  the minimal area $S$ in terms of a
given contour $C$, defined by quark and antiquark paths $z_{\mu}(t)$ and
$\bar{z_{\mu}}(t)$.

Any surface can be parametrized by the Nambu-Goto form
\begin{equation}
S= \int^{T}_{0} d \tau \int^{1}_{0} d\beta [\dot{w}^2_{\mu} w'^{2}_{\mu}
-(\dot{w}_{\mu}w'_{\mu})^2]^{1/2}
\end{equation}
where $w_{\mu}(\tau,
\beta)$ are the coordinates of the string world surface, and $$\dot{w}_{\mu}
= \frac{\partial w_{\mu}}{\partial \tau}, w'_{\mu}= \frac{\partial
w_{\mu}}{\partial \beta}$$.

We use the approximation  that the minimal surface for given paths $z_{\mu}
(\tau)$,\\
$\bar{z}_{\mu}(\tau)$ is determined by eq.(17) with $w_{\mu}$  given
 by straight lines, connecting points $z_{\mu}(\tau)$ and
$\bar{z}_{\mu}(\tau)$ with the same $\tau$, i.e.  \begin{equation}
w_{\mu}(\tau,\beta) = z_{\mu}(\tau)\cdot
\beta + \bar{z}_{\mu}(\tau )(1-\beta)\;\; , \;\;\;0 \preceq \beta \preceq 1
\end{equation}
where $\tau$ is defined for both trajectories as
\begin{equation}
\tau =\frac{t}{s}T = \frac{\bar{t}}{\bar{s}}T
\end{equation}
and $T$ enters boundary condition.
We have no proof that eq. (18) indeed fits the minimal area $S$  in
all cases; we note that this approximation is valid  in two limiting
cases which are of special interest below: in the case
$l=0$ one can exploit  the
flat dynamics of quarks and in the  limit $l\rightarrow
\infty$ corresponding to the dynamics of the  string with typical
trajectories of the double helycoid type, for which the minimal area indeed
is formed by the straight-lines.  In what  follows we shall use eqs.
(14,17,18) to express $<W(C)>$ in terms of quark coordinates.

We introduce a compact notation
\begin{equation}
\xi  \equiv \{ \tau, \beta\}\;\;,\;\;\; g_{ab}(\xi) \equiv \partial_a
w_{\mu}\partial_{b} w^{\mu},\;\;\; a,b = \tau,\beta \end{equation} and
 obtain \begin{equation} S= \int d^2 \xi  \sqrt{det g}.  \end{equation}

\section{The reduction of  four-dimensional
\newline
dynamics to the three-dimensional one}

Before the insertion of the concrete form of $<W(C)>$ let us consider the
integration over $z_{0}(t), \bar {z}_{0}(t)$ and its physical interpretation
(in the meson rest system). These integrations
(additional with respect to the nonrelativistic case)
are related to  the following phenomena.
  First, the relativistic quantum theory in
general is the theory with unconserved number of particles.
Even in the case of a free particle the backtracking of the time-component
which corresponds to the creation of additional pairs at the intermediate
stage  leads to the proper relativistic structure of Green's function.
Second the fact, that the interaction time is smeared out, generates a need
for the integration over the relative time of particles in the process even
in the valence quark approximation.

We suggest the following approach to the problem. At first
step we separate out from the initial sum over all paths
 the class of the trajectories without backtracking of
time-components, satisfying the condition (in the rest system)

\begin{equation}
\frac{dz_{0}(\gamma_{1})}{d\gamma_{1}} > 0~,
{}~~\frac{d\bar{z}_{0}(\gamma_{2})} {d\gamma_{2}} > 0
\end{equation}
or
rewriting it in the explicitlly Lorentz-invariant form

\begin{equation}
(P^{\mu}~dz_{\mu}(\gamma_{1})/ d\gamma_{1}) > 0~,~~(P^{\mu}
d\bar{z}_{\mu}(\gamma_{2})/ d\gamma_{2}) > 0
\end{equation}
where $P_{\mu}$
is the total 4-momentum of the meson.

This constraint has to be imposed, because for the trajectories with such
backtracking the minimal surface which appears in $<W(C)>$
obviously can not be properly approximated
by our straight line anzatz. This  procedure corresponds to
the usual valence quarks approximation and in the language
of the time ordered diagrams the trajectories with backtracking
corresponds to the production of $q\bar{q}$ - mesonic states.

 In Appendix A it is shown, that for the
trajectories without time-back\-tracking the variables $z_{0}(\gamma_{1})~,
{}~~\bar{z}_{0}(\gamma_{2})$ can be transformed into the new nondynamical ones
$\mu_{1}(\tau)~,~~\mu_{2}(\tau)$ without derivative terms
by using the parametrization for which
\begin{equation}
z_{\mu}=(\tau~, \vec{z})~~, \bar{z}_{\mu}=(\tau~,~\vec{\bar{z}})
\end{equation}

  The
integration over these new variables for the spinless  quarks
is performed effectively by steepest decent method and
leads only to
the modification of the initial kinetic terms $K, \bar{K}~(8)$ in the
quark Green's function in an external field (7). The final action
becomes independent of the time-components of $z_{\mu}, \bar{z}_{\mu}$.

\begin{equation}
G(x\bar{x} / y\bar{y})=\int D\mu_{1}~D\mu_{2}~D\vec{z}~D\vec{\bar{z}}~
exp[-K'-\bar{K}'] <W(C)>_{A}
\end{equation}
where
\begin{equation}
K'+\bar{K}'=\int\limits^{T}_{0} \frac{d\tau}{2}
[(\frac{m^{2}_{1}}{\mu_1(\tau)} +\mu_{1}(\tau) \{ 1+\dot{\vec{z}}^2(\tau)
\})+ (\frac{m^{2}_{2}}{\mu_{2}(\tau)} +\mu_{2}(\tau) \{1+\dot{\vec{z}}^{2}
(\tau) \} )]
\end{equation}
We have
 introduced the new proper time parameter $\tau~,~~ 0\leq
\tau \leq T$~, and

\begin{equation}
\mu_{1}(\tau)=\frac{T}{2s}\dot{z}_{0}(\tau)~,~~
\mu_{2}(\tau)=\frac{T}{2\bar{s}}
\dot{\bar{z}}_0(\tau)
\end{equation}
with the dot standing for the derivative over $\tau$. It should be stressed
that
the substitution of eq.(24) into $<W(C)>$
is implied.
This condition is equivalent to
\begin{equation}
R_0(\tau)=\tau,\;\;\; r_0(\tau) =0
\end{equation}
which corresponds to the usual instant plane Hamiltonian
dynamics in the c.m. system. We also observe in (28) that $\tau$ plays the
 role of the
time for the meson -- a common time for both quarks.  We shall prove, that
in the case of the pure straight-line string the condition
 $$(Pr)=0$$
directly follows from the dynamics of the string.  This result supports
the validity of our approximation of neglecting
(in the straight-line approximation for the minimal
surface) of backward in time trajectories in the rest system. It is
important to stress, that  this assumption can be not valid  for  the
calculation of quantities other than the mass spectrum and for
interactions different from the
confining string-like interaction of quarks, which we have
formulated.

Therefore we have shown that the starting 4D dynamics can be reduced
to the 3D one. And it should be emphasized that in general this is a
nonlocal 3-dimensional dynamics due to the double independent integration
in $W(C)$
over $\vec{z}(\tau)$ and $\vec{\bar{z}}(\tau)$, which bindes together
$\vec{z}(\tau)$ and $\vec{\bar{z}}(\tau)$ with different arguments
$\tau_{1}$ and $\tau_{2}$.

In addition we have two additional integrations over $D\mu_1$, and
$D\mu_2$ which replace integrations over time  components
$z_0,\bar{z}_0$, and one common evolution parameter  in the
action which can be identified with the c.m. time. The physical meaning of
$\mu_1,\mu_2$ can be most easily clarified  imposing an external e.m.
field, e.g. in calculating the magnetic moment. Then it can be shown,
that $\mu_i$ enter magnetic moment as the constituent mass of a quark. The
same happens with spin-dependent forces (to be published). Therefore we
shall call $\mu_i$ the dynamical mass of the quark $i$. In the approximation
when $\mu$ does not depend on $\tau$, this dynamical mass has been
introduced in [2,7]. As we shall see below this approximation works
reasonably well -- accuracy in the determination of mass is around 5\%
( see e.g. Table 4 of [2]).

\section{Gaussian representation for the effective action of quarks and the
 string.}

Combining the results of the previous section we obtain the
 total effective action.

\begin{equation}
A\equiv K'+\bar{K'} + \sigma_0\int^{T}_{0} d\tau \int^{1}_{0} d\beta \sqrt{det
g}
\end{equation}
 This action is similar to the one considered in the papers [15]
 , where the straight-line string without transverse excitations has been
 considered. However in our case also kinetic terms of quarks $K'+\bar{K}'$
 are present and the condition that the ends off the string move with the
velocity of light is not
 possible.

  A direct procedure of quantization of (29) is difficult due to the square
root
  term and we use the auxiliary fields approach [3]  to get rid of it.
  In the  Appendix B we give a detailed derivation of this procedure , while
here we present
  the final gaussian representation of this action

\begin{eqnarray}
A &=& \int^T_0 d\tau \int^1_0 d\beta \{ \frac{1}{2}(\frac {m^2_1}{\mu_1(\tau)}
+
\frac {m^2_2}{\mu_2(\tau)})+ \frac{1}{2} \mu_{+}(\tau) \dot{R}^2 + \frac
{1}{2}{\tilde \mu}(\tau) \dot{r}^2 + \\ \nonumber
&+& \frac{1}{2\tilde{\nu}}[\dot{w}^2 + (\sigma \tilde {\nu})^2
r^2 - 2\eta (\dot{w}r)+ \eta^2 r^2]\}
\end{eqnarray}
where
$$\mu_{+}(\tau)= \mu_1(\tau)+\mu_2(\tau)~,~~~\tilde{\mu}(\tau) =
\frac{\mu_1(\tau)\cdot \mu_2(\tau)}
{\mu_1(\tau)+\mu_2(\tau)}$$
and  the condition (24) for $z_{\mu}(\tau), \bar{z}_{\mu}(\tau)$ is
assumed.  Since only extremal values of $\mu_1(\tau)$ and $\mu_2(\tau)$
contributes one can intgrate in our symmetric case over restricted class of
functions $\mu_1(\tau)=\mu_2(\tau)=\mu(\tau)$.
Here $\tilde{\nu}(\tau,\beta) \geq 0$ and $\eta(\tau,\beta)$ are
two auxiliary fields, which should be integrated out in the full path
integral representation for $G$:
\begin{equation}
G = \int DR Dr D\tilde{\nu} D\eta D\mu e^{-A}
\end{equation}

In order to prove the  equivalence of the dynamics governed by the actions (29)
and (30)
it is enough to  show that after proper integration over
$\tilde{\nu} (\tau, \beta),\eta (\tau, \beta)$ one returns from eq. (30)
 back to the
initial one (29).

Actually after the gaussian integration over $\eta (\tau, \beta)$ we obtain
instead of eq. (30) the following expression for G
\begin{equation}
G= \int
DR Dr D\mu exp [ -K'-\bar{K}'] D\tilde{\nu}\times
\end{equation}
$$exp [-\int^T_0 d\tau \int^1_0 d\beta \frac{1}{2} [\frac{\dot{w}^2
w'^2-(\dot{w}w')^2} {(r^2\tilde{\nu})}+ \sigma^2 (\tilde{\nu} r^2)]]$$
where we restore the notation $$w_{\mu}' = r_{\mu}$$.

For the integral over $\tilde{\nu}$, as it is shown in Appendix A,
the explicit measure can be constructed and the effective
action is determined only by the extremum value of $\tilde{\nu}$.

\begin{equation}
\sigma^2 r^4 \tilde{\nu}^2_0 = (\dot{w}^2 w'^2 -(\dot{w} w')^2)
\end{equation}
Inserting this expression for $\tilde{\nu}$ into the eq.(32)
one recovers  our starting action (29).

We emphasize that the integration over $\tilde{\nu}$ and $\eta$
effectively amounts to the replacement of them by their extremum values.

 The resulting action (30) is quadratic in $R_{\mu}, r_{\mu}$ and can be
 conveniently rewritten as (here $\mu_1 = \mu_2 = \mu, m_1 = m_2 = m$)
\begin{equation}
A= \int^T_0 d\tau[\frac{m^2}{\mu} +\frac{1}{2}\{a_1 \dot{R}^2 + 2a_2
(\dot{R} \dot{r})- 2c_1(\dot{R}r)- 2c_2(\dot{r}r)+
a_3 \dot{r}^2 + a_4 r^2\}]
\end{equation}
where we have used
 $w_{\mu}=R_{\mu}+(\beta-1/2)r_{\mu}$
 to express $\dot{w}$ in terms of $\dot{R},\dot{r}$.
We have introduced the following notations
\begin{eqnarray}
\nonumber
a_1 = \int^1_0 d\beta(2{\mu} + \nu), \;\;\; a_3 = \int^1_0 d\beta(\frac{\mu}{2}
+
(\beta -\frac{1}{2})^2\nu) \\
a_2 = \int^1_0 d\beta(\beta- \frac{1}{2})\nu, \;\;\; a_4 = \int^1_0
d\beta(\frac{\sigma^2}{\nu}
 + \eta^2 \nu)\\
\nonumber
c_1 = \int^1_0 d\beta \eta \nu, \;\;\; c_2 = \int^1_0 \eta \nu(\beta -
\frac{1}{2})
d \beta
\end{eqnarray}
and $$\nu(\tau,\beta) = \frac{1}{\tilde{\nu}(\tau,\beta)}$$ plays the
role of the dynamical energy density for the string.  Since only
$\dot{R}_{\mu}$ enters the action then it is convenient to integrate over
$D\dot{R}$~~ taking into account boundary conditions $R_{\mu} (T)-
R_{\mu}(0) = u_{\mu}T$, i.e.
\begin{equation}
\int DR \rightarrow \int
D\dot{R} \int^{i \infty}_{-i \infty} d^4
\lambda exp( \lambda_{\mu} \int^T_0
(\dot{R}_{\mu} - u_{\mu}) d\tau)
 \end{equation}

We shall systematically disregard  in what follows the preexponential factors
and
shall be interested
 only in the effective action in the
exponent. In this way integrating out $D\dot{R}$ as in (36) we obtain
\begin{equation}
G\sim \int Dr D\nu D\eta D\mu d^4 \lambda \exp (-\tilde{A})
\end{equation}
where
$$ \tilde{A} = \frac{1}{2} \int^{T}_0 d\tau \{ 2(\lambda u) +
\frac{2m^2}{\mu} + \frac{1}{a_1}[(a_3 a_1 - a_2^2)\dot{r}^2
 + 2(c_1 a_2 - c_2 a_1)
(r \dot{r}) $$
\begin{equation}
+ (a_4 a_1 - c_1^2) r^2 + 2a_2(\lambda \dot{r}) -
2c_1 (\lambda r) - \lambda^2\}
\end{equation}

 For generality we preserve the explicitly  Lorentz- invariant  form of the
action,
 with the conditions $R_0 =1, r_0 =0$  being implied.  It  apparently amounts
to
 the replacement in eq.(36) of an  ordinary integration $d\lambda_0$ by the
functional one $D\lambda_0(\tau)$.  The expression (38) for the action
$\tilde{A}$ formes the basis of our further calculations.

\section{A quantization of the pure straight-line string}

In this section the case of the pure straight-line string will be studied.
It appeares that in this approximation to the full theory one doesn't need
to use the conditions $(Pr)
=0$. We shall show that in this case the
reparametrization symmetry (B.3) dynamically induces this condition.
We shall also obtain in this section the spectrum of this particular system.

In Appendix B it is shown that for the straight line string without
kinetic terms of quarks at the ends the effective action (38) can be reduced
to the following one

\begin{eqnarray}
G=\int~DR_{\mu}~Dr_{\mu}~D\eta(\tau,\beta) ~D\tilde{\nu}(\beta)\\ \nonumber
exp[-\int\limits^{T}_{0}d\tau \int\limits^{1}_{0}d\beta
\frac{1}{2\tilde{\nu}}
[\dot{w}^{2}+((\sigma\tilde{\nu})^{2}+\eta^{2})r^{2}-2\eta(\dot{w}r)] ]
\end{eqnarray}
with $\tilde{\nu}(\beta)$ being independent on $\tau$.

As well as in the previous section it is convenient to introduce instead
of $\tilde{\nu}(\beta)$ a new variable,

\begin{equation}
\nu(\beta)=1/ \tilde{\nu}(\beta)
\end{equation}
which plays the role of an effective energy density of the string.

Let us first consider an integration over $D\eta(\tau, \beta)$.
The functions $\eta(\tau, \beta)$ enter the action as an integral over
$\beta$ with functions $\nu(\beta)$.  Only the extremal values of the
 function $\nu(\beta)$, which are even under the exchange $\beta
 -\frac{1}{2} \rightarrow - (\beta - \frac{1}{2})$ ( due to the symmetry
 under the permutation of the ends of the string) contributes to the action.
 So we shall integrate only over the  class of functions $\nu (\zeta)(\zeta
= \beta - \frac{1}{2})$, which are even functions of $\zeta$.  It is
convenient to decompose  the functions $\eta(\tau, \beta)$ into orthogonal
polinomials $P_n(\beta)$ with the weight $\nu(\beta)$ \begin{equation}
\eta(\tau, \beta) = \sum_n P_n(\beta) k_n(\tau)
\end{equation}
\begin{equation}
\int^1_0 d\beta \nu(\beta) P_n(\beta) P_m(\beta) = \delta_{mn}
\end{equation}

Taking into account the symmetry of $\nu(\beta)$, one can easily obtain
expressions for
$P_0(\beta)$
\begin{eqnarray}
\nonumber
P_0(\beta) = N_0,\;\;\;\  N^2_0 = (\int^1_0 \nu(\beta) d \beta) ^{-1}\\
P_1(\beta) = (\beta - 1/2) N_1, \;\;\; N_1^2 = (\int^1_0 \nu (\beta)
(\beta - 1/2)^2 d\beta)^{-1}
\end{eqnarray}

The  action (39) can be written in terms of $N_i$ in the following form
\begin{eqnarray}
\nonumber
 S&=& \frac{1}{2} \int^T_0 d\tau [
\frac{1}{N_1}\dot{r}^2 + \sigma^2\int \frac{d\beta} {\nu(\beta)} r^2 + r^2
\sum^{\infty}_{n=1} k^2_n(\tau) - 2N_0 k_0(\tau) i (\lambda r)-\\
&-&\frac{2}{N_1}(\dot{r}r) k_1(\tau)+\lambda^2 N^2_0 +2i(\lambda u)]
\end{eqnarray}
with $\lambda$ being replaced by  $i\lambda$.

The function $k_0(\tau)$ enters only in the 4th term in this expression
and an integration over $Dk_0(\tau)=\prod^{N}_{i=1}dk_0(\tau_i)(N\rightarrow
\infty)$ gives the factor proportional to a product of $\delta$-functions;
\begin{equation}
\prod^{\infty}_{i=1} \delta (\lambda r(\tau_i))
\end{equation}

Thus only the components of $r_{\mu}$ transverse to the direction of
$\lambda_{\mu}$, which plays the role of the total momentum $P_{\mu}$ should
be taken into account and we have
\begin{equation}
\int D^4r \delta
(\lambda r) exp [-S] \rightarrow \int D^3r~ exp[-S]
\end{equation}

After the integration (46) the dependence of $S$ on $\lambda$ has a simple
form \begin{equation} T(\frac{\lambda^2}{a_1}+2i(\lambda u)) \end{equation}
and the integral over $\lambda$  is saturated in the case $T\rightarrow\infty$
we are interested in by an extremum value
\begin{equation}
\lambda_{\mu} = i\frac{1}{N_0^2}  u_{\mu}= P_{\mu}
\end{equation}

Integration over $Dk_n$ with $n\geq 1$
 leads effectively to the following expression
\begin{equation}
\int d\nu(\beta)D^4r\delta(ru) exp [-\frac{1}{2} \int^{T}_{0} d\tau
[\frac{1}{N^2_0} +
(\dot{r}^2 -\frac{(r\dot{r})}{r^2})\frac{1}{N^2_1}+
\sigma^2\int\frac{d\beta}{\nu(\beta)} r^2 ]
\end{equation}
It is important that we have obtained the following constraints
\begin{equation}
(ru)\sim(rP)=0
\end{equation}
\begin{equation}
(rp)=0
\end{equation}
where $P_{\mu}$ is the total momentum of the string and
\begin{equation}
p_{\mu} =(\dot{r}_{\mu}-\frac{(r\dot{r})}{r^{2}}r_{\mu})\frac{1}{N^2_1}
\end{equation}
is the relative momentum of the string.

The second constraint (51) means that only transverse components of $p$
enter the action. The same constraints appear in the canonical
quantization of the straight-line string [14] .

Consider the rest system of the meson $u_{\mu}=(1,\vec{0})$ and transform the
expressions from the
Euclidean to the Minkowski space
\begin{equation}
d\tau_E\rightarrow id\tau_M
\end{equation}

It follows from  eq. (49) that the hamiltonian of the problem is
\begin{equation}
H(\vec{p},\vec{r}) =\frac{1}{2} \{N^2_1 \frac{\hat{L}^2}{\vec{r}^2}+
\sigma^{2} \int \frac{d\beta}{\nu}\vec{r}^2+\frac{1}{N^2_0}\}
\end{equation}
where $\hat{\vec{L}}=(\vec{r}\times \vec{p})$ is the operator of the angular
momentum.

This hamiltonian does not contain the radial part of the kinetic term so that
the field $\vec{r}^2$ is not a dynamical one. Noticing that the integral over
$\vec{r}^{2}$ has the form (A.13), one
concludes that in the effective action only the extremum of $\vec{r}^{2}$
contributes. So that after solving the equation of motion (for a fixed
value of orbital momentum)

\begin{equation}
-\frac{l(l+1)}{\vec{r}^4}N^2_1 +
 \sigma^2 \int \frac{d\beta}{\nu}=0
\end{equation}
we arrive at the final expression for the hamiltonian  as follows
\begin{equation}
H(\nu, l)=\frac{1}{2} \frac{1}{N_0^2} +
\sigma  \sqrt{\int \frac{d\beta}{\nu}N^2_1} \sqrt{l(l+1)}
\end{equation}

The function  $\nu(\beta)$ has the form
which gives the minimum of this hamiltonian
(for details see ref.[13])

\begin{equation}
\nu_{l}(\beta) = (\frac {8 \sqrt{l(l+1)}\sigma}{\pi})^{1/2}
\frac{1}{\sqrt{1-4(\beta - \frac{1}{2})^2}}
\end{equation}
This solution corresponds to the spectrum of the hamiltonian
\begin{equation}
E^2_l =M^2_l = 2\pi \sigma \sqrt{l(l+1)}
\end{equation}
which agrees with the result obtained for the straight-line string in the
canonical
formalism [14].

Expression (57) could be physically interpreted if one notices that a
parameter

\begin{equation}
v(\beta)=2(\beta-1/2)
\end{equation}
plays the role of the velocity of the corresponding elementary piece of
the string. In this way formula (57) can be rewritten into more familiar one

\begin{equation}
\nu_{l}(\beta)=\frac{\rho_{l}}{\sqrt{1-v^2(\beta)}}
\end{equation}
where $\rho_{l}=(\frac{8\sigma\sqrt{l(l+1)}}{\pi})^{1/2}$ corresponds to the
effective mass density of the string in the rest frame.

To conclude this section let us give the physical interpretation of the
 constraint
(50), which shows that (in the rest system) the  relative time is unimportant
 for the
dynamics of the pure straight-line string.
The given realization of  the string doesn't include internal
interaction between neighbouring points of the string, so that there is no
physical exchange along the string.
Because of this fact  there is no  need for  the introduction of relative
time dynamics.

\section{The general case of the QCD string with quarks}

 This is the central part of our paper. In this section we shall derive the
 effective hamiltonian for the general case of the straight-line
  QCD-string with quarks and shall find the spectrum of the problem.
 As it has been done
in the previous section, we integrate over $\eta(\tau,\beta)$, expanding it in
orthogonal polinomials $P_n(\beta)$
with weight $\nu(\tau,\beta)$
 After this integration  and integration over $\vec{\lambda}$
 in the c.m. system $\vec{u}=0$ we obtain for the equal mass case
 $(\mu_1(\tau) = \mu_2(\tau)= \mu ((\tau))$

\begin{eqnarray}
S = \int\limits^{T}_0 d\tau (\frac{m^2}{\mu(\tau)} + \mu(\tau) + \frac{1}{2}
\{\frac {\mu(\tau)\dot{\vec{r}}^2}{2} +\\ \nonumber
\int(\beta - 1/2)^2 \nu d\beta \frac{(\dot{\vec{r}} \times \vec{r})^2}
{\vec{r}^2} + \int \frac{\sigma^2 d\beta}{\nu} \vec{r}^2 + \int \nu d\beta \}
\end{eqnarray}
which will form the basis of our further discussions.

It should be stressed, that the absence of the dependence on the relative
time $r_0(\tau)$ ( the so-called instantaneous interaction) doesn't mean
 of course that this effective action is induced by the  interaction between
 quarks and string during an infinitely small time  interval.

 Actually, we have managed to tranform the integration over $r_0 (\tau)$
 into the one over $\mu(\tau)$. And it is the peculiar property of the
 considered interaction that after this transformation one can (under the
 approximation discussed above) arrive at a local three-dimensional
 dynamics.

 We shall now consider  first the case of heavy masses and then arbitrary
 masses with $l$ increasing from $l=0 $ to $l=\infty$. \\

\underline{The limit of nonrelativistic potential dynamics.}\\

This is the case when $m \gg \sqrt{\sigma}$, so that
\begin{equation}
m \sim <\mu> \gg <\nu>
\end{equation}

An extremal equation for $\mu(\tau)$
\begin{equation}
\frac{m^2}{\mu^2(\tau)} = 1 + \frac{\dot{\vec{r}}^2}{4}
\end{equation}
gives after taking into account  a nonrelativistic condition
\begin{equation}
\dot{\vec{r}}^2 \ll 1
\end{equation}
the simple solution in the leading order
\begin{equation}
\mu(\tau) = m
\end{equation}
{}From the extremal condition for $\nu(\tau,\beta)$
\begin{equation}
\frac{\sigma^2}{\nu^2(\tau~,~\beta)} \vec{r}^2 = 1 + (\beta - 1/2)^2
\frac{(\dot{\vec{r}} \times \vec{r})^2}{\vec{r}^2}
\end{equation}
one obtains in the leading order
\begin{equation}
\nu(\beta~,~\tau) = \sigma\sqrt{\vec{r}^2}
\end{equation}
and finally the action in the same approximation is
\begin{equation}
S = \int\limits_0^T d\tau[2m +\frac{m}{4}\dot{\vec{r}}^2 +
\sigma\sqrt{\vec{r}^2} ]
\end{equation}
as one can expect from the very beginning.\\

\underline{The general case. The transition from the potential dynamics}\\
\underline{to the string dynamics.}\\

We start with the case $l=0$ and arbitrary masses $m$
\begin{equation}
L^2 \sim (\dot{\vec{r}} \times \vec{r})^2 = 0
\end{equation}
One obtains from (61) with the help of (66)

\begin{equation}
S = \int\limits_0^T d\tau [\frac{m^2}{\mu(\tau)} + \mu(\tau)
+\frac{\mu(\tau)}{4}
\dot{\vec{r}}^2 + \sigma\mid\vec{r}\mid ]
\end{equation}

In the Minkowski space-time the action (70) yields the Hamiltonian

\begin{equation}
H = \mu(\tau) + \frac{\vec{p}^2 + m^2}{\mu(\tau)} + \sigma\mid\vec{r}\mid
\end{equation}
and the extremal condition for $\mu(\tau)$ is

\begin{equation}
\mu(\tau) = \sqrt{\vec{p}^2 + m^2}
\end{equation}
Therefore we arrive at the following Hamiltonian in the case $l=0$
\begin{equation}
H(\vec{p},\vec{r}) = 2\sqrt{\vec{p}^2+ m^2} + \sigma\mid\vec{r}\mid
\end{equation}
where  $\vec{p}^2 = (\vec{p}\vec{r})^2/\vec{r}^2 \equiv p^2_r$

This expression is widely used in the context of the so-called "relativistic
quark model" [11]. As was discussed in [2,7] the eigenvalues of (73)
differ only a little from the approximate version of this Hamiltonian
 used by one of the authors.
There eq.(71) has been used with $\mu$ independent
of $\tau$. As a consequence, the procedure was to find first eigenvalues of
(71) $E(\mu)$ as a function of $\mu$, and then to minimize $E(\mu)$ with
respect to $\mu$, i.e.  to find $\mu=\mu_0$ from $\frac{dE}{d\mu}=0$, and
to calculate $E(\mu=\mu_0)$. As can be seen from Table 4 of [2], the
eigenvalues
$\varepsilon_n$ of (73) and $E_n(\mu_0)$ differ at most by 5\% for lowest
states, while calculations of $E(\mu_0)$ are much easier to do, especially
for many-quark and gluon states.

Let us discuss how the potential $\sigma\mid \vec{r}\mid$ obtained in the
rest frame is transformed under Lorentz boosts. One should keep in mind that it
is
induced by the area law of Willson's loop, which is a Lorentz scalar. Therefore
it is not difficult to verify that this potential is a Lorentz scalar also and
in
arbitrary frame it can be represented as
$$\sigma (r^2_{\mu}-\frac{(Pr)^2}{P^2})^{1/2}$$
where $P$ is the total 4-momentum of the hadron .

In the case of small  values of $l$, the
string contribution to the kinetic part of the action (and to the total orbital
momentum)

\begin{equation}
\int(\beta-1/2)^2\nu(\beta)
d\beta\frac{(\dot{\vec{r}}\times\vec{r})^2}{\vec{r}^2}
\end{equation}
can be treated  as a perturbation of the Hamiltonian (73) (see [2] for
a discussion).
If the Hamiltonian (73) (valid only for low $l$ and strictly speaking for
$l=0$ ) would be used for calculation
 of the spectrum for arbitrary values of $l$, one could obtain [11]
\begin{equation}
 M^2 =2\pi \sigma (2n_r+ \frac{\lambda(n_r)}{\pi} l + \delta(n_r,l))
\end{equation}
 where $\lambda(n_r) \rightarrow 4 $ when $l\gg n_r$ and $ \delta (n_r,l)$
 is the  small correction for all values of $n_r$ and $l$. This formula
 gives a good approximation of the spectrum for not large $l$.
   For the Regge trajectory at large $l$  one would get

$$ M^2 = 8\sigma l $$
However this approximation  gives  $\approx 25\%$ deflection from the
correct value, as we shall demonstrate below.

Now we obtain the Hamiltonian of the system for  arbitrary $l$.
Separating longitudinal and transverse
components of $\dot{\vec{r}}$ with respect to $\vec{r}$, we obtain

\begin{equation}
\dot{\vec{r}}^2 = \frac{1}{\vec{r}^2} \{ (\dot{\vec{r}} \vec{r})^2 +
(\dot{\vec{r}}\times \vec{r})^2 \}
\end{equation}
and the kinetic part of the action (61) can be written as

\begin{equation}
\frac{1}{2}(\frac{\mu(\tau)}{2}\frac{(\dot{\vec{r}}\vec{r})^2}{\vec{r}^2}
+ (\frac{\mu(\tau)}{2} + \int\limits_{0}^{1}(\beta - 1/2)^{2}
\nu(\beta,\tau) d\beta) \cdot~\frac{(\dot{\vec{r}}\times
\vec{r})^2}{\vec{r}^2} ) \end{equation} For the longitudinal and transverse
components of the momentum one gets respectively

\begin{equation}
p^2_r \equiv \frac{(\vec{p}\vec{r})^2}{\vec{r}^2} = (\frac{\mu}{2})^2
\frac{(\dot{\vec{r}}\vec{r})^2}{\vec{r}^2}
\end{equation}

\begin{equation}
p^2_T \equiv \frac{(\vec{p} \times \vec{r})^2}{\vec{r}^2} = (\frac{\mu}{2} +
\int \limits_{0}^{1} (\beta-1/2)^2 \nu(\beta,\tau) d\beta)^2
\frac{(\dot{\vec{r}}\times \vec{r})^2}{\vec{r}^2}
\end{equation}
The standard derivation of $H$ from the action in terms of these
components yields in the Minkowski space-time

\begin{eqnarray}
H(p,r,\nu,\mu) =
\frac{1}{2}(\frac{(p^2_r+m^2)}{\mu(\tau)/2} +2\mu(\tau) +\\ \nonumber
\frac{\hat{L}^2/\vec{r}^2}{(\frac{\mu}{2}+ \int (\beta-1/2)^2 \nu d\beta)} +
\int \frac{\sigma^2 d\beta}{\nu} \vec{r}^2 + \int \nu d\beta)
\end{eqnarray}
where
\begin{equation}
\hat{L}^2 = (\vec{p} \times \vec{r})^2 = p^2_T \vec{r}^2
\end{equation}

This is the resulting Hamiltonian for the QCD straight-line string with quarks.

Postponing consideration of this general expression to future papers
let us now concentrate on the transition of the dynamics from the potential
case
for small $l$ with the Hamiltonian (73) to the case of large $l$, which we
call the  string dynamics.

In the limit $l \rightarrow \infty$ one can expand the potential part of
the Hamiltonian (80)
\begin{equation}
\frac{1}{2}(\frac{\hat{L}^2/\vec{r}^2}{(\mu /2+ \int (\beta-1/2)^2
\nu d\beta)} + \int \frac {\sigma^2}{\nu} d\beta \vec{r}^2)
\end{equation}
around the extremum in $\mid{\vec{r}}\mid$

\begin{equation}
r^2_l = (l(l+1)/(\mu /2 + \int (\beta-1/2)^2 \nu d\beta) \cdot
\int \frac{\sigma^2}{\nu} d\beta)^{1/2}
\end{equation}

Keeping only quadratic terms in $(r - r_l)$ one gets instead of (80)
\begin{eqnarray}
H=1/2 \{\frac{p^2_r+m^2}{\mu(\tau)/2} +2\mu(\tau) + \int \nu d\beta +\\
\nonumber
2(\frac{\sigma^2(l(l+1)) \int \frac {d\beta}{\nu} }{\mu/2+
\int (\beta-1/2)^2 \nu d\beta})^{1/2} + 4 \int \frac {\sigma^2}{\nu} d\beta
(\mid{\vec{r}}\mid - r_l)^2 \}
\end{eqnarray}
Let us show that  in the limit $l \rightarrow \infty$ the dynamical masses
$\mu$
and $\nu$ satisfy the condition

\begin{equation}
< \mu > \ll < \nu >
\end{equation}
In this case  expanding eq.(84) in $\mu / \nu$, we shall  obtain the string
dynamics regime and the leading Regge trajectory of the Nambu-Goto form
\begin{equation}
M^2_l \rightarrow 2\pi \sigma l
\end{equation}

 Now we  prove that the alternative regime
\begin{equation}
< \mu >\;\; >\;\; < \nu >
\end{equation}
yields larger mass values
and therefore it  is energetically disfavoured. From eq. (84)
 one can
 simply
estimate  that the regime
\begin{equation}
< \mu > \gg < \nu >
\end{equation}
is not possible for  $l \rightarrow \infty$ at all  and we are left with a
possibility
\begin{equation}
< \mu > \approx < \nu >
\end{equation}
which corresponds  to the situation when quarks at the ends carry
the  fraction of the total energy (orbital momentum)
 comparable
with the string contribution.

In this case one can treat the term $\int (\beta-1/2)^2 \nu d\beta$
perturbatively. Starting with the value $r_l$
\begin{equation}
r^2_l = (\frac {l(l+1)}{\mu}\sigma^{2} \int \frac{d\beta}{\nu})^{1/2}
\end{equation}
and exploiting a numerically small coefficient
$$ \int (\beta-1/2)^2 d\beta = \frac{1}{12}$$
One returns to the Hamiltonian (73) which leads
 to the trajectory
\begin{equation}
M^2 \rightarrow 8 \sigma l
\end{equation}
with a larger mass for a given $l$  than in eq.(86). This  demonstrates
 that the relativistic  potential regime
 is disfavoured, as compared to the "string" one.

We come back now to the string dynamics regime (85) where quarks at the
ends carry only small part of the total energy (orbital momentum) of the
hadron. In the leading approximation neglecting the dynamics of the
longitudinal components one gets

\begin{equation}
S^{(0)}_L = \int\limits_{0}^{T} (\frac{1}{2} \int \nu d\beta +
(\frac{\sigma^{2} l(l+1) \int \frac{d\beta}{\nu}}{\int(\beta-1/2)^2\nu
d\beta})^{1/2})d\tau
\end{equation}
The extremal value of $\nu(\tau,\beta) = \nu_l^{(0)}$ is $\tau$--independent,
and we
recover the case of the pure string (56)
\begin{equation}
E^{(0)}_l=\frac{1}{2} \int\limits_{0}^{1} \nu(\beta) d\beta +
(\frac {\sigma^2l(l+1)\int \frac{d\beta}{\nu(\beta)}}{\int(\beta-1/2)^2
\nu(\beta) d\beta})^{1/2}
\end{equation}
and hence
\begin{equation}
\nu^{(0)}_l(\beta) = (\frac{8\sigma(l(l+1))^{1/2}}{\pi})^{1/2}
(1 - 4(\beta -1/2)^2)^{-1/2}
\end{equation}

$$(E^{(0)}_l)^2 \rightarrow 2\pi \sigma l$$

The corrections to the eq.(92) are considered in Appendix C. It should
be stressed that while the qualitative dependence  of the dynamical mass
$<\mu>$ on $l$ is reasonable
\begin{eqnarray}
<\mu>\sim l^{\alpha} \;\;\; \alpha>0 \;\;\; l\rightarrow \infty\\
\nonumber
<\mu> /<\nu> \rightarrow 0 \;\;\;\;\;l\rightarrow \infty
\end{eqnarray}
the quantitative results are out of the
accuracy of the straight-line approximation.\\

\section{Conclusion}

We have applied in this paper the path integral method to quantize the
quark-antiquark system interacting nonperturbatively. We have argued that
the latter leads to the appearence of a minimal string between the quarks
at large distances, $R \gg T_g$, where $T_g\simeq 0.2 fm$ is the correlation
length in
the vacuum.

As a result our starting Lagrangian consists of kinetic terms for
quarks (including the relative time term) and
 the string part. Using the method of auxiliary fields one gets an effective
action quadratic in coordinates and its derivatives. The auxiliary fields
are participating in the final action and the integration measure for them is
found.
We have shown that for the spectrum the integration amounts to taking the local
extremum
of the action in the values of auxiliary fields.

One of the important problems dealt with in the paper is the question of the
relative time. We have shown that this question is  resolved for the case
of the pure straight-line  string due to the reparametrization invariance
 of the action.

It leads to the constraint $(Pr)=0$, which corresponds to the  condition
$R_0=0$ in the rest system. We argued that for a string  with quarks the
same constraint should be used in order to make the straight-line
approximation for the surface selfconsistent. It corresponds to the choice
of quark trajectories without backward motion in time and defines the
valence quark approximation. For this class of trajectories it is possible
to eliminate the relative time using the reparametrization invariance of the
string term resulting in a Hamiltonian with variable dynamical masses
$\mu_i(\tau)$. This leads to the transformation of the initial quadratic
kinetic
terms to the square - root type ones.

Thus we have obtained the three dimensional dynamics with dynamical
variables connected to the dynamical masses of quarks $(\mu_{i}(\tau))$ and
to the effective dynamical string mass density $\nu(\tau,\beta)$.

The spectrum depends on the relative role of the quark and string degrees of
freedom (d.o.f.) which leads to the emergence of two different dynamical
regimes.
 We have  obtained it analytically in two limiting
cases. (i) States with the total angular momentum $l=0$. Here in the kinetic
part of the action only the quark degrees of freedom contribute while the
string
provides only the inert part, namely, the exactly linear potential. The
Hamiltonian appears to be equal to the so-called "relativistic quark model"
[11]
 one and eigenvalues coincide within the accuracy of our model with those of
the
 proper-time Hamiltonian [2,7] where $\mu_1$ do not depend on $\tau$.
This potential-like relativistic
regime (when the string carries only a small part of the total orbital momentum
$l$ of the hadron) is valid up to moderate values of $l$ smoothly joining the
string-like regime at large $l$.
  (ii) For the  states
with $l \gg 1$
 the string degrees of freedom  dominate. The string carries not only energy,
 but orbital momentum
 and cannot be reduced
to the potential term only. The effective Hamiltonian for radial degrees
of freedom is derived and we show that it yields
only small  corrections to the string
levels.

It is gratifying to note, that the Regge slope of our trajectories is
asymptotically $(2\pi\sigma)^{-1}$, the usual string slope, and it differs
by $\approx 25 \%$  from what one would get from
the potential regime (i). The latter slope is $(8\sigma)^{-1}$, the
same as was obtained long ago in the so-called relativistic quark potential
model [11] and in [2,7]. It should be stressed, that the  string slope is
energetically favorable as compared to the potential one.

The case of the pure straight-line string (without quarks at its ends) has been
treated
in detail in Section 6. Results here
coincide with those obtained by the canonical quantization method  [14]. This
case corresponds to the limit (ii) of $l\rightarrow \infty$ of the previous
one.

One should note also, that our results are in an agreement  with
numerical quantization of the same quark-string system, done in ref.[21].
There the instantaneous dynamics has been assumed from the beginning and
quasiclassical quantization was performed numerically. The authors [21] also
have observed two limiting regimes; that of $l=0$ and $l\gg 1$.\\
\newpage
\setcounter{equation}{0}
\renewcommand{\theequation}{A.\arabic{equation}}

{\bf Appendix  A}\\

{\bf A particle Green's function without backward motion}\\

{}~~~~~We shall obtain below the Green's function of a particle in the external
field under the condition $\frac{d z_{0}(\gamma)}{d\gamma}>0$ for its motion.
This condition implies that the only trajectories taken into account
are those without additional pair creation, corresponding to the backward
in the time motion $\frac{dz_{0}}{d\gamma} < 0$.

To illustrate the idea we first consider the case of a free particle
subject to the same condition

\begin{equation}
\frac{dz_{0}(\gamma)}{d\gamma} > 0
\end{equation}

The standard form of the free particle Green's function in the
Minkowski space

\begin{eqnarray}
G(x,y)=<x\mid(-\partial^{2}+m^{2})\mid y>\sim
\int\frac{d\tilde{s}}{\tilde{s}^2}
Dz_{\mu}(\gamma)\times\\ \nonumber
\times~~ exp[-i \int\limits^{1}_{0} \frac{d\gamma}{2}(\frac{m^2}{\tilde{s}}-
\dot{z}^2_{\mu}\tilde{s})]=\int_0^{+\infty} d\tilde{s}
[exp [\frac{-i}{2}(\frac{m^2}{\tilde{s}}-
(x-y)^2 \tilde{s})]]
\end{eqnarray}
contains the summation over all trajectories with any sign of
$\frac{dz_{0}}{d\gamma}$. In the momentum space $G$ has the form

\begin{equation}
G(p_1~,~p_2)\sim \delta^{4}(p_1-p_2)\frac{1}{p^2_1 - m^2}
\end{equation}

One can separate in (A.2) the sum over a class of trajectories without backward
motion (A.1) as follows. For each trajectory of this class one makes in a
unique way
the change
of integration parameter $d\gamma$ by $\frac{dz_{0}}{\dot{z}_{0}}$

\begin{equation}
d\gamma=\frac{dz_{0}}{\dot{z}_0}
\end{equation}
and the exponent in (A.2) can be rewritten as

\begin{eqnarray}
exp[-i \int\limits^{1}_{0}
\frac{d\gamma}{2}(\frac{m^2}{\tilde{s}}-\dot{z}^{2}_{\mu}\tilde{s})]
\rightarrow [-i
\int\limits^{T}_{0}\frac{dz_{0}}{2}(\frac{m^2}{\tilde{s}\dot{z}_{0}}+
\tilde{s}\dot{z}_{0}(1-\dot{\vec{z}}^{2}(z_{0})))]=\\ \nonumber
=~exp[-i\int\limits^{T}_{0} d\tau(\frac{m^2}{\tilde{s}\dot{z}_{0}(\tau)}+
\tilde{s}\dot{z}_0(\tau)(1-\dot{\vec{z}}^{2}(\tau)))]
\end{eqnarray}
with

\begin{equation}
T=x_{0}-y_{0}=z_{0}(T)-z_{0}(0)
\end{equation}

In the class of trajectories with property (A.1) the transition from the
integration over $d\tilde{s} Dz_{0}$ to $d\mu(\tau)$ with

\begin{equation}
\mu(\tau)=\tilde{s}\dot{z}_{0}
\end{equation}
has a nonsingular Jacobean, well known in the string theory [12]

\begin{equation}
D\mu^2(\tau)\sim exp[-i\frac{const}{\epsilon}\int\limits^{t}_{0}
 \sqrt{\mu^2(\tau)}d\tau]
d\tilde{s} Dz_{0}(\tau)
\end{equation}
where $\frac{1}{\epsilon}\sim\Lambda$ is the ultraviolet cut-off parameter. We
note
that $\tilde{s}$ plays role of a collective coordinate for a set
$\{\mu(\tau)\}$, since from (A.7) one has

\begin{equation}
\tilde{s}=\frac{1}{T}\int\limits^{T}_{0}d\tau \mu(\tau)
\end{equation}

Combining (A.4-A.8) one obtains for the Green's function $\tilde{G}$ in the
class of trajectories (A.1) the following expression

\begin{equation}
\tilde{G}=\int D\vec{z}(\tau)D\mu^2(\tau)exp[-i\int\limits^{T}_{0}
\frac{d\tau}{2}(\frac{m^2}{\mu(\tau)}+\mu(\tau)(1-\dot{\vec{z}}^{2}(\tau)))]
\end{equation}
where a proper rescaling of the mass and $\vec{z}(\tau)$ is made. Note that
in this representation the extremum value of $\mu(\tau)$ is

\begin{equation}
\mu(\tau)=m\sqrt{1-\dot{\vec{z}}^{2}(\tau)}
\end{equation}
which makes it difficult to treat the case $m=0$ and trajectories with
$\dot{\vec{z}}^{2} > 1$.

Hence it is convenient to use  the canonical, path integral form where one gets

\begin{equation}
\tilde{G}=\int D\vec{z}(\tau)D\vec{p}(\tau)d\mu(\tau)
exp[i\int\limits^{T}_{0}(\vec{p}\dot{\vec{z}}-\frac{1}{2}
\{ \frac {\vec{p}^{2}+m^2}{\mu(\tau)}+\mu(\tau) \} )d\tau]
\end{equation}

It is important to define the correct integration measure. Having in mind that
the
following equality holds true

\begin{equation}
\int\limits^{+\infty}_{0}\frac{dt}{\sqrt t} exp[-\frac{a}{2}(t+1/t)]=
2(\frac{\pi}{2a})^{1/2}e^{-a}
\end{equation}
for $\mid \arg a\mid \leq \pi/2$, we are to choose (taking into account  eq.
(A.8))

\begin{equation}
D\mu(\tau)\sim\sqcap_{i}\frac{d\mu(\tau_{i})}{\mu^{3/2}(\tau_{i})},
\end{equation}

After the integration over $\mu(\tau)$ one obtains for $\tilde{G}$

\begin{equation}
\tilde{G}=\int D\vec{z}(\tau)D\vec{p}\;
exp[i\int\limits^{T}_{0}d\tau(\vec{p}\dot{\vec{z}}-\sqrt{\vec{p}^2+m^2})]
\end{equation}
where the exponent is given by the extremum value of $\mu(\tau)$:

\begin{equation}
\mu(\tau)=\sqrt{\vec{p}^2(\tau)+m^2}
\end{equation}

The expression (A.15) is the usual canonical representation
 for the quantum mechanical Green function with the Hamiltonian
$$H=\sqrt{\vec{p}^2+m^2}$$

In the momentum representation one has instead of eq.(A.3)

\begin{equation}
\tilde{G}(p_1\;,\;p_2)\sim \delta^3(\vec{p}_1 - \vec{p}_2)
(E_1-\sqrt{\vec{p}^2_1
+m^2})^{-1}
\end{equation}

Thus the contribution of the trajectories without backward motion is
equivalent (for a free particle) to the separation of the positive frequency
part
 from the relativistic propagator.

We take now the case of a particle in the external field $A_{\mu}$. In an
analogous way from the standard form of the Green's function in the external
field

\begin{equation}
G(x,y \mid A_{\mu})\sim \int \frac{d\tilde{s}}{\tilde{s}^2}Dz_{\mu}
exp[-i\int\limits^{1}_{0} d\gamma(\frac{1}{2}(\frac{m^2}{\tilde{s}}-
\tilde{s} \dot{z}_{\mu}^2)-g\dot{z}_{\mu}A^{\mu})]
\end{equation}
one obtains in the class of trajectories (A.1) the following
Green function in external
field

\begin{equation}
\tilde{G}_{inv}(x,y\mid A_{\mu})\sim \int D{\mu}(\gamma)Dz_{\mu}(\gamma)
exp[-i\int\limits^{T}_{0} d\tau(\frac{1}{2}(\frac{m^2}{{\mu}(\gamma)}+
\mu(\gamma)(1-\dot{\vec{z}}^2))-gA_{\mu}\dot{z}^{\mu})]
\end{equation}

In the same way as it has been done for the free particle to transform
eq.(A.10) into eq.(A.15) , one can prove that the expression (A.19) leads
to usual relativistic Hamiltonian of the particle in external field
\begin{equation}
H(\vec{p}~, \vec{x}~~A_{\mu})= -g A_{0}(\vec{x}, \tau )+\sqrt{(\vec{p} +
g\vec{A}(\vec{x},\tau))^2 +m^2}
\end{equation}

\setcounter{equation}{0}
\renewcommand{\theequation}{B.\arabic{equation}}

{\bf Appendix B}\\

{\bf The auxiliary field formalism}\\

To develop a procedure of quantization of (29) we use the auxiliary fields
formalism, as is usually done in the string theory [3,12].

Let us rewrite (29) as

\begin{equation}
G=\int Dr~DR~D\mu~~exp[-K^{\prime}-\bar{K}^{\prime}]~Dh_{ab}
exp[-\sigma_{0}\int \sqrt {h}
d^2\xi]~\cdot \\
\end{equation}
$$~\cdot \delta(\partial_{a}w_{\mu}\partial_{b}
w^{\mu}-h_{ab}(\xi))= \int Dr~DR~Dh_{ab}\int\limits^{+i
\infty}_{-i\infty}D\lambda^{ab} exp[-\sigma_{0}\int \sqrt {h} d^2\xi]
{}~\cdot$$\\
$$ \cdot exp[+\int \sqrt{h}\lambda^{ab}h_{ab} d^2
\xi] exp [-\int \sqrt{h} \lambda^{ab} \partial_{a}w_{\mu}\partial_{b}w^{\mu}
d^{2}\xi] exp[-K-\bar{K}]$$
where
\begin{equation}
d^{2}\xi=d\gamma d\beta~,~~ \xi_{1}=\gamma~,~~ \xi_2=\beta~,
{}~~h\equiv deth.
\end{equation}

In the pure string case when the kinetic terms are absent the action (B.1)
is invariant under reparametrization  $\gamma \rightarrow f(\gamma,\beta)$
\begin{eqnarray}
w_{\mu}(\gamma,\beta)\rightarrow w_{\mu}(f(\gamma,\beta),\beta)~,
{}~h_{11}(\gamma,\beta)\rightarrow (\frac{\partial
f}{\partial\gamma})^2h_{11}(f,\beta),\\ \nonumber
h_{12}(\gamma,\beta)\rightarrow (\frac{\partial
f}{\partial\gamma})h_{12}(f,\beta)~,
{}~~h_{22}(\gamma,\beta)\rightarrow h_{22}(f,\beta)
\end{eqnarray}
with the function $f(\gamma,\beta)$ satisfying the conditions

\begin{equation}
f(0,\beta)=0~~, f(1,\beta)=1~~, \frac{\partial f(\gamma,\beta)}
{\partial\gamma} > 0
\end{equation}

It appears to be convenient to decompose [12]
\begin{equation}
\lambda^{ab}(\xi)=\alpha(\xi)h^{ab}(\xi)+f^{ab}(\xi)
\end{equation}
with
\begin{equation}
f^{ab}h_{ab}=0~~, h^{ab}\equiv(h^{-1})^{ab}
\end{equation}

The integral (B.1) takes the form
\begin{eqnarray}
G=\int Dr~DR~Dh_{ab}exp[-K^{\prime}-\bar{K}^{\prime}]\int
D\alpha(\xi)Df^{ab}(\xi)\\ \nonumber
exp[-\int(\sigma_{0}-2\alpha(\xi))\sqrt {h} d^2\xi] \cdot\\ \nonumber
\cdot exp [-\int \sqrt{h}((\alpha(\xi)h^{ab}+f^{ab})
\partial_{a}w_{\mu}\partial_{b}w^{\mu})
d^{2}\xi]
\end{eqnarray}

We will show now that in the continuum limit $\alpha(\xi)$ and $f^{ab}(\xi)$
can be replaced by their mean values

\begin{equation}
< \alpha(\xi) > \rightarrow \bar{\alpha}~~,
< f^{ab}(\xi) >\rightarrow 0
\end{equation}

Equation (B.8) reflects the fact, that $\alpha(\xi)$  is a scalar, while
$f^{ab}(\xi)$ is a traceless tenzor.

One can  write $\alpha(\xi)$ in the form

\begin{equation}
\alpha(\xi)=<\alpha(\xi)>(1+b(\xi))
\end{equation}

Since $\alpha(\xi)$ is a scalar, one can expand

\begin{equation}
<\alpha(\xi)>=\bar{\alpha}+c_1 R_1(\xi)+\ldots
\end{equation}
where $R_1(\xi)$ is the scalar curvature for the metrics $h_{ab}$;
$\bar{\alpha}$  is some constant with a magnitude much larger (as we
will show below) than a characteristic value of $R_1(\xi)$, so that one
can neglect all terms except the first one.

To simplify the problem we neglect the dependence of $h_{ab}$ on $\xi$ and
put $h_{ab}(\xi)=\delta_{ab}$. In this case it is sufficient to consider
a model problem, corresponding to Eq. (B.7)

\begin{equation}
\int D\alpha(\gamma)~DR(\gamma)~Dr(\gamma)~exp(-\int\limits^{\prime}_{0}
d\gamma \alpha(\gamma) \dot{R}^{2}) exp(-\int\limits^{\prime}_{0}
\alpha(\gamma)(\dot{r}^{2}+r^2)d\gamma) \end{equation}

We need the effective action of $\alpha(\gamma)$. The Gaussian integration
over $DR(\gamma)$ yields

\begin{equation}
\int \frac{dq}{2\pi}b(q)b(-q) \int\frac{dk}{2\pi}~~\frac{[k(q-k)]^2}
{k^2(q-k)^2}
\end{equation}

The integral over $dk$ is linearly divergent and should be cut off at
$k \sim \Lambda\rightarrow\infty$. Therefore its singularities in the
variable $q$ lie at a distance $\sim\Lambda$

\begin{equation}
B(q^2)=\int \frac{dk}{2\pi}\frac{[k(q-k)]^2}{k^2(q-k)^2} \sim\Lambda(1+const
(\frac
{q}{\Lambda})^2)
\end{equation}
and the induced action

\begin{equation}
W_R \sim \Lambda \int b^2(\gamma)d\gamma
\end{equation}
suppresses fluctuations of $b(\gamma)$.

The $r(\gamma)$ contribution to the effective action can be obtained with
the help of transformation

\begin{equation}
r(\gamma)=f(\gamma)\int\limits^{\gamma}_{0} \frac{\dot{z}(\gamma^{\prime})
d\gamma^{\prime}}{f(\gamma^{\prime})}
\end{equation}
where $f^{\prime \prime}-f=0~, f(0)=f(1)$~~, and we obtain

\begin{equation}
S_r=\int\limits^{1}_{0} d\gamma \alpha(\gamma)(\dot{r}^{2}+r^2)=
\int\limits^{1}_{0} \alpha(\gamma)\dot{z}^{2}(\gamma)+
\int\limits^{1}_{0} d\gamma\{ \alpha(\gamma)\frac{d}{d\gamma}
(f^{\prime}f \int\limits^{\gamma}_{0}\frac{\dot{z}}{f}d\gamma^{\prime})\}
\end{equation}

Note that the Jacobian

\begin{equation}
\frac{Dz}{Dr}=(\frac{f(1)}{f(0)})^{1/2}
\end{equation}
does not depend on $\alpha(\gamma)$.

As in the previous case one obtains a contribution to the effective action
from the first term on the r.h.s. of (A.16)

\begin{equation}
W_r \sim \Lambda \int b^2(\gamma)d\gamma
\end{equation}
which damps again fluctuations of $f(\gamma)$, and therefore
the second term in expression (A.16) is unimportant.

Thus the field $\alpha(\gamma)$ is connected in fact with two free fields
$R(\gamma)$ and $z(\gamma)$. In this case $\bar{\alpha} \sim
\Lambda\rightarrow\infty$
as shown in [12].

In the same way one can show that the effective action for the fields
$f_{ab}(\xi)$
also damps fluctuations of fields

\begin{equation}
W\sim \Lambda \int f^{2} d\gamma
\end{equation}

Therefore we have justified the equality (B.8).

After that we obtain the following expression for $G$:

\begin{eqnarray}
G=\int Dr~DR~D\mu_{1}~D\mu_{2}~Dh_{ab}~[-K^{\prime}-\bar{K}^{\prime}] \cdot\\
\nonumber
\cdot exp[-(\sigma_{0}-2 \bar{\alpha}) \int \sqrt{h}d^2{\xi}]
\cdot exp [-\bar{\alpha}\int \sqrt{h} h^{ab} \partial_{a}
w_{\mu}\partial_{b}w^{\mu} d^{2}\xi]
\end{eqnarray}
with the new action which is quadratic in $w_{\mu}$ and contains the new
auxiliary fields $h_{ab}$.

Let us  first consider the case of the pure string. The invariance
(B.3) makes it convenient to introduce the new variables
$\tilde{\nu}(\beta),~ f(\xi),~ \eta(\xi)$,~ separa\-ting out the collective
mode $\tilde{\nu}(\beta)$ and the field $f(\gamma,\beta)$, satisfying
conditions (B.4)

\begin{equation}
\hbar\equiv \frac{h}{h^2_{22}}=(T\sigma \tilde{\nu}(\beta))^2
(\frac{\partial f (\gamma,\beta)}{\partial\gamma})^2
\end{equation}
and making a simple rescaling of $\hbar_{12}$.

\begin{equation}
\hbar_{12}(\xi)\equiv \frac{h_{12}}{h_{22}}=(\frac{\partial f (\gamma,\beta)}
{\partial\gamma})(T\eta(\gamma,\beta))
\end{equation}
where $T$ enters  the boundary condition. Taking into account the fact,
that

\begin{equation}
Dh_{11}~Dh_{22}~Dh_{12} = D\hbar~D\hbar_{12}~h_{22}^{2}~Dh_{22}
\end{equation}

and using the well known  formula [12]
\begin{equation}
D\hbar^2 \sim exp[- \frac{const}{\epsilon}\int\sqrt{h}
d^2{\xi}]D\tilde{\nu}(\beta)
Df(\gamma,\beta)
\end{equation}
where $1/ \epsilon \sim \Lambda$ is the ultraviolet cut-off scale, we
arrive at the following expression after change of the integration over
$d\gamma$ by $Tdf(\gamma,\beta) \equiv d\tau$
$$
 G=\int Df Dh_{22} Dr DR D\eta
D\tilde{\nu}(\beta)\cdot$$
 $$\cdot exp[-(\sigma_0-2\bar{\alpha}+
\frac{const}{\varepsilon})\sigma \int h_{22} (\xi)\tilde{\nu}(\beta) d\tau
d\beta]\cdot$$
\begin{equation}
\cdot exp[-\int^{T}_0 d\tau \int^1_0 d\beta
\frac{1}{2\tilde{\nu}} ((\frac{\partial w}{\partial\tau})^2 - 2\eta
(\frac{\partial w}{\partial \tau} r) +((\tilde{\nu}\sigma)^2 +
\eta^2)r^2)]
 \end{equation}
 where trivial rescaling  $z,\bar{z}\rightarrow
(\frac{\sigma}{2\bar{\alpha}})^{1/2}z,
(\frac{\sigma}{2\bar{\alpha}})^{1/2}\bar{z}$ together with a proper
renormalization of $m_0,
s$ in $K,\bar{K}$ has been done.

At first we notice that the action doesn't depend on $f(\gamma,\beta)$
which reflects the invariance (B.23). So that the integral over
$Df(\tau,\beta)$ can be factored out and it is equal to the volume of the
reparametrization group.

In the standart way [12] we have introduced the physical quantity $\sigma$,
which entered our expression (B.21)

\begin{equation}
\sigma^2=\bar{\alpha}(\sigma_0 - 2\bar{\alpha}+\frac{const}{\epsilon})
\end{equation}

In the general case (B.1) the invariance (B.3)
 is lost so that  we are left
with a dependence of $\tilde{\nu}$ on $\gamma, \beta$ and have to restrict
ourself instead of (B.21), (B.22) by a simple rescaling

\begin{equation}
\hbar\equiv \frac{h}{h^2_{22}}=(T\sigma \tilde{\nu}(\tau,\beta))^2
\end{equation}

\begin{equation}
\hbar_{12}\equiv \frac{h_{12}}{h_{22}}=
(T\eta(\gamma,\beta))
\end{equation}

After that in the same way we arrive at the expression for $G$

\begin{eqnarray}
G =\int Dr~DR~D\mu~Dh_{22} D\eta D\tilde{\nu}(\tau,\beta)~\\ \nonumber
\times exp[-K^{\prime}-\bar{K}^{\prime}]exp[-(\sigma_{0}-2 \bar{\alpha})
\sigma\int h_{22}(\xi) \tilde{\nu}(\tau, \beta)d\tau d\beta]\\ \nonumber
\times exp [-\int\limits^{T}_{0} d\tau \int\limits^{1}_{0}d\beta
\frac{1}{2\tilde{\nu}}((\frac{\partial w}{\partial\tau})^2 -
2\eta(\frac{\partial w}{\partial\tau}r)+((\tilde{\nu}\sigma)^2+\eta^2)r^2)]
\end{eqnarray}
which differs from (B.45) by three points; (i) changing factor
$(\sigma_{0}-2\bar{\alpha} +\frac{const}{\epsilon})$ by
$(\sigma_{0}-2\bar{\alpha})$ (ii) an explicit dependence of
$\tilde{\nu}(\tau,\beta)$ on $\tau$ (iii) by the presence of kinetic terms.

Finally after gaussian integration over $h_{22}\geq 0$ we obtain for the case
of the pure string

\begin{equation}
G=\int DR_{\mu}~Dr_{\mu}~D\tilde{\nu}(\tau)d\eta(\tau,\beta)
exp[-A_{str}]
\end{equation}

where
\begin{equation}
A_{str}=\int\limits^{T}_{0} d\tau \int\limits^{1}_{0} d\beta
\frac{1}{2\tilde{\nu}}
[\dot{w}^2 + (\sigma \tilde{\nu})^2 r^2 - 2\eta(\dot{w}r)+\eta^2 r^2]
\end{equation}
and in the general case (B.1)
\begin{equation}
G=\int DR_{\mu}~Dr_{\mu}~D\mu~D\tilde{\nu}(\tau,\beta)d\eta(\tau,\beta)
exp[-A_{str}-K^{\prime}-\bar{K}^{\prime}]
\end{equation}
\\

\setcounter{equation}{0}
\renewcommand{\theequation}{C.\arabic{equation}}

{\bf Appendix C}\\

In this Appendix we derive the effective Hamiltonian for the longitudinal
excitations, calculate the corrections to the string result (85), and
consider the behaviour of the dynamical mass $\mu$ as a function of $l$.

It is easy to prove that after
taking into account the longitudinal dynamics one obtains a small correction
to (93) of the order of
$$\mu^{-1/2}l^{-3/4}$$
Since $\mu$ is weakly growing with $l$, as we shall see below, one concludes
 that at $l \rightarrow \infty$ the field $\nu(\tau,\beta)$ separates and is
governed by its own purely string dynamics, only weakly perturbed by the
dynamics of $r^2(\tau)$ and $\mu(\tau)$. The latter are "living" in the
external
field $\nu(\beta)$.

We now compute the longitudinal contribution to the energy of the system to the
leading order in $(1/l)$. To this end one can use the nonperturbed value
$\nu^{(0)}_l$ (93) and make an expansion in (83) as follows.

\begin{eqnarray}
2(\frac{\sigma^2 l(l+1) \int \frac{d\beta}{\nu^{(0)}}}{\frac{\mu}{2}+ \int
(\beta-1/2)^2 \nu^{(0)} d\beta})^{1/2} = 2(\frac{\sigma^2 l(l+1) \int
\frac{d\beta}{\nu^{(0)}}}{\int (\beta-1/2)^2 \nu^{(0)} d\beta})^{1/2} + \\
\nonumber
 +(-1)\frac{\mu}{2}\frac{(\sigma^2 l(l+1)\int\frac{d\beta}{\nu^{(0)}})^{1/2}}
{(\int(\beta-1/2)^2 \nu^{(0)} d\beta)^{3/2}}
 +3/4(\frac{\mu}{2})^2
\frac{(\sigma^2 l(l+1)\int \frac{d\beta}{\nu^{(0)}})^{1/2}}
{(\int(\beta-1/2)^2 \nu^{(0)} d\beta)^{5/2}}
\end{eqnarray}

Insertion of $\nu^{(0)}(\beta)$ from (93) into (C.1) yields
\begin{equation}
(2\pi\sigma(l(l+1))^{1/2})^{1/2}-2{\mu}+ \frac{3\cdot\sqrt{2}}
{\sqrt{\pi}} (\mu)^2(\sigma(l(l+1))^{1/2})^{-1/2}
\end{equation}
and finally one gets
\begin{eqnarray}
H = \frac{1}{2} [ \frac{p^2_r+m^2}{\mu/2}+2(2\pi\sigma)^{1/2}(l(l+1))^{1/4}+\\
\nonumber
+\frac{\mu^2}{4} \frac {12\sqrt{2}}{\sqrt{\pi}(\sigma^2l(l+1))^{1/4}}+
\frac{\sigma^{3/2}\pi^{3/2}(r - r_0)^2}{2^{3/2} (l(l+1))^{1/4}} ]=
\\
\nonumber
=E^{(0)}_L + H^{(r)}
\end{eqnarray}

where $H^{(r)}$ is an effective Hamiltonian for the radial excitations of the
hadron

After insertion in this expression of the extremal value of $\mu$
\begin{eqnarray}
\mu = (\frac{p^2_r+m^2}{3})^{1/3}(\frac{\pi\sigma l}{2})^{1/6}
\end{eqnarray}
we finally obtain for the effective radial Hamiltonial
\begin{eqnarray}
H^{(r)}=\frac{1}{2}(3^{4/3}(\frac{2}{\pi\sigma l})^{1/6} (p^2_r+m^2)^{2/3}+
(\frac{\pi \sigma}{2})^{3/2} l^{-1/2} (r-r_0)^2)
\end{eqnarray}
where the substitution $(l(l+1))^{1/2} \rightarrow l$ has been made in the
limit $l\rightarrow \infty$.

Let us consider the case of massless current  quarks
\begin{eqnarray}
m=0
\end{eqnarray}
Introducing instead of $(r-r_0)$ a new dimensionless variable $x$
\begin{eqnarray}
(r-r_0)= 3^{2/5} l^{1/10}(\frac{\pi \sigma}{2})^{1/2} x
\end{eqnarray}
we represent eigenvalues of the Hamiltonian (C.5) in the following way
\begin{eqnarray}
E^{(r)}_{l,n_r}= l^{-3/10}(\frac{\pi\sigma}{2})^{1/2} 3^{4/5}a(n_r)
\end{eqnarray}
where $a(n_r)$ is the eigenvalues of the new dimensionless Hamiltonian
\begin{eqnarray}
\tilde{H}^{(r)} = \frac{1}{2}((-\frac{d^2}{dx^2})^{2/3} +x^2)
\end{eqnarray}
In order to obtain the approximate value $a(n_r)$ we consider eq. (C.3) for
the restricted class of functions $\mu$ independent on $\tau$. Such
procedure in general gives accuracy about 5\% for low lying states. In this
way one can easily get
\begin{eqnarray}
 a(n_r)=2^{-1/5}\cdot 3^{-3/5}
\frac{5}{4} (n_r+\frac{1}{2})^{4/5}
\end{eqnarray}

Substituting this expression into eq. (C.8) we arrive at the final
expression for the total energy $E_{l,n_r}$ of the hadron and have
for the mass squared
\begin{eqnarray}
M^2_{l,n_r}= 2\pi\sigma (l+ const
l^{1/5}(n_r + 1/2)^{4/5})
\end{eqnarray}
 which is slightly different in the case $l\rightarrow \infty $ from
the pure string result (85).


\end{document}